# Secure Transmission of Sensitive data using multiple channels

Ahmed A. Belal, Ph.D.
Department of computer science and automatic control
Faculty of Engineering
University of Alexandria
Alexandria, Egypt.
aabelal@hotmail.com

Abdelhamid S. Abdelhamid, M.Sc.
Department of computer science and automatic control
Faculty of Engineering
University of Alexandria
Alexandria, Egypt.
asaladin@acm.org

## Abstract

*A new scheme for transmitting sensitive data is proposed, the proposed scheme depends on partitioning the output of a block encryption module using the Chinese Remainder Theorem among a set of channels. The purpose of using the Chinese Remainder Theorem is to hide the cipher text in order to increase the difficulty of attacking the cipher. The theory, implementation and the security of this scheme are described in this paper.*

## 1. Introduction

As the need to protect sensitive data against different threats [2] increases the study of cryptography becomes of greater importance [1].

Encryption is the cryptographic primitive mostly used in protecting the secrecy of the data. Block encryption can be viewed as a mathematical transformation between the set of source blocks of size $N_S$ bits and the set of blocks of size $N_C$ bits through a mapping function and a parameter $K$ that is called the encryption key. The form of any block Encryption is $C = E_K(S)$ for some $S \in \{0,1\}^{N_S}$ and $C \in \{0,1\}^{N_C}$. The inverse transform known as the decryption is $S = D_K(D)$.

The Chinese remainder theorem CRT [5] states that if $q_0, q_1...q_{k-1}$ are k pair wise relatively prime positive integers and $a_0, a_1...a_{k-1}$ are positive integers then there exists exactly one integer $a$ where $0 \leq a < q$ for $q = \prod_{i=0}^{k-1} q_i$ such that $a = a_i \bmod q_i \ for \ 0 \leq i < k$.

The integers $q_0, q_1...q_{k-1}$ are called the moduli while the integers $a_0, a_1...a_{k-1}$ are called the residues. The CRT has well known applications in both secret sharing and error correcting codes [6].

In this work, block encryption and the Chinese remainder theorem will be combined to produce a scheme for transmitting sensitive data over multiple channels.

## 2. Overview of the proposed scheme

The proposed scheme assumes the existence of $A$ transmission channels between the sender and the receiver parties from which $S$ Channels are chosen using some selection criteria.

The original message or plain-text is divided into units that are referred to as super blocks of N bits. The super blocks are encrypted using an encryption module that operates using a block encryption scheme $E_K(B)$ of block size $N_B$ bits where N is equal to $LN_B$ for an integer L. The encryption is performed using an appropriate mode of operation such as CBC [2]. Each super block is treated as an N bit integer.

A set of $S$ relatively prime moduli $q_0, q_1...q_{S-1}$ are selected such that $2^N \leq q_0 q_1..q_{S-1}$ then the N bit integer CRT remainders with respect to the selected $S$ moduli $r_0, r_1...r_{S-1}$ are calculated then sent over the selected $S$ channels. The Selected S channels are used to transmit the remainders while the rest of the channels carry some irrelevant data in order to decrease the ability of the adversary to determine the used channels.

At the receiver side, the inverse of the CRT is applied to get the original N-bit cipher super block, and then a decryption module $D_K(B)$ is used to get the original N bit plain-text.

The application the CRT to the output of the encryption module aims to hide the cipher from the adversary in order to prevent the adversary from taking benefit of any property of the cipher.

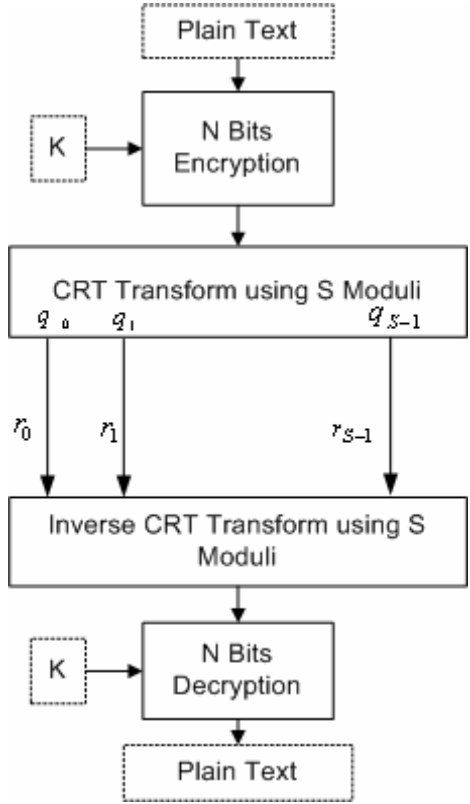

**Figure 1 Proposed Scheme Block Diagram**

## 2.1 Selection of the Channels

The selection of the $S$ out of the A available channels should be done in a way that prevents the adversary from telling which channels carry relevant data and which ones do not. The proposed selection criterion is based on selecting pseudorandom numbers in the range $[0, A-1]$. The generator is run on both sides of communication $S$ times to select the $S$ channels.

## 2.2 Selection of the moduli

It is important to determine the range of values for the moduli used in the Chinese Remainder transform. Assuming that, the size of the super block is N bits, and that it can be treated as an N-bit integer. The mod operation can be considered as a mapping between the N-bit data space and the residue space accordingly any super block $b$ can be mapped to a remainder $r$ such that $0 \leq r < q$ and $r = b \bmod q$ where $q$ is the selected modulus.

Assuming a random variable B that denotes the possible super block integer value and a random variable R that denotes the possible values of the residue, the joint probability between the two random variables can be represented as $P(B=b/R=r) = q/2^N$.

In order to decrease this probability we either increase the super block size N or reduce the value of the modulus $q$. In order to decrease the value of all moduli the size of all of them can be chosen to be around the value $\lceil N/S \rceil$.

## 2.3 Number of channels

The maximum number of channels used in the proposed scheme was estimated as $S_{\max} = O(N/\log N)$. This estimation is based on the fact that the maximum number of relatively prime numbers that can be selected less than or equal to a given integer x is $O(x/\log x)$ and the assumption that the moduli will be selected of length $\lceil N/S \rceil$.

## 2.4 The super block Size

The super block size N is selected as a multiple of the block size $N_B$ of the encryption scheme used such that $N = LN_B$. The selection of the multiplier L is governed by some factors

1. The loss resulting from representing the remainders of the Chinese remainder theorem on fixed word length machines. For example if AES is used as the encryption method and 10 channels are used. If the encryption module super block size was taken to be 128 bits i.e. L=1, each channel will carry 12.8 bits or practically 13 bits so 2 bytes are used for the representation of that number which leads to 18.75% loss in the bandwidth. If the super block size was taken to be 512 bits i.e. L=4, each channel will carry 51.2 bits or practically 52 bits so 7 bytes will be used leading to only 7.14% loss in bandwidth.

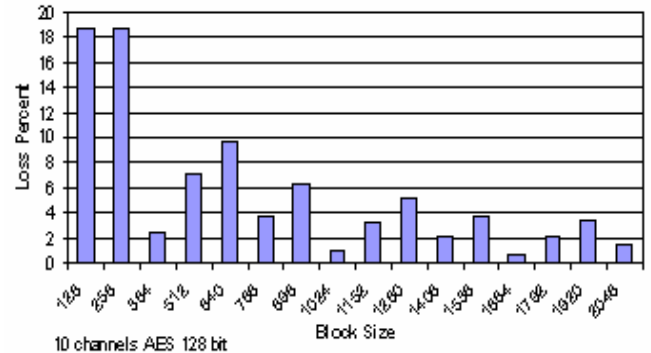

**Figure 2 bandwidth Loss**

2. The network algorithms used to manage the queuing of data. For example in some implementation of the TCP/IP the Nagle algorithm is used in order to reduce the network overhead. The application of that algorithm results in the buffering of data till a certain threshold is reached then this data is sent as one packet. The use of larger remainders will result in faster filling of buffer and therefore increasing the throughput.

3. As the value N increases the length of the moduli selected increases. This will increase the effort needed by the adversary to guess the moduli

## 3. The detailed scheme

The proposed scheme is composed of two phases:

1. The setup phase, during this phase the setup for the scheme will be done. This phase is done once or every a relatively long interval.
2. The session phase, during this phase the actual transmission is done between the two parties. The session phase starts as soon as the setup phase ends and is repeated while the setup configuration is valid.

*Scheme*()
{

  *SetupPhase*();
  // IsValid represents outside conditions
  *while*(*IsValid*)
    *Session*();

}

**Figure 3 Proposed Scheme Flow**

### 3.1 The setup phase

In order to setup the environment for the proposed scheme the following steps should be done:

1. Selection of the Encryption method, its parameters and mode of operation. For example if we decide to use AES in the CBC mode, both the Encryption key and the initial vector should be selected. An important parameter to be set is the super block size $N = LN_B$.
2. Establishment of the set of available channels.
3. Selection of the number of channels used in the communications S, this number should be selected such that it is possible to select at least S relatively prime integers to be used in the CRT.
4. Selection of the S channels out of the A available channels. This is done through the use of a suitable pseudorandom number generator [3], [4]. The main condition of the random number generator is to produce a full cycle of values in the range $[0, A-1]$. The output of the pseudorandom generator is taken to represent channels IDs $[0, A-1]$ so the generator is run S times to generate the S channels ids selected for the session.
5. Selection of the moduli. Two ways for using these moduli are proposed
   a. To generate a number of moduli equal to the number of channels A, such that each channel is assigned a static modulus. It should be kept in mind that from the generated A moduli any S moduli can be used to represent the output of the encryption module using the CRT. This method will be referred to as the static assignment of moduli.
   b. To generate only S moduli sufficient to represent the output of the Encryption module using CRT and on each session the S moduli are assigned to the S selected channels. This method will be referred to as the dynamic assignment of moduli.

The output of the setup phase represents the key for the proposed scheme and so some sort of key exchange protocol should be used in order to exchange the output of the setup phase.

### 3.2 The session phase

The session phase refers to the steps that should be done on each transmission. It involves activities on both sides of the transmission.

In the following discussion the two parties will be referred to as the sender and the receiver. Every session starts as soon as the previous session ends. The operations Input Wait, channels Input Wait, and channel Read will be assumed to be blocking operations.

The session does not start as soon as the data arrives in order to keep the synchronization between the sender and the receiver parties. It should be noted that the first session starts as soon as the setup phase ends. In the following, the two different implementations for the Sender and the Receiver will be discussed.

*Sender Session*

The Sender has a message M that should be delivered to the receiver. The protocol steps used by the sender are:

1. Select channels step, this step selects S channels from the Available Channels using a pseudorandom number generator. After selecting the channels the moduli are assigned to the corresponding channels.
2. Wait for input step, this step waits until there are data available for sending. This operation is a blocking step. When the data is available steps three to six will be repeated until the input is consumed.
3. N bits of data are read from the input stream. If the amount of remaining data is less than N, the data will be padded as appropriate by the application.
4. The N-bit super block will be encrypted by the encryption module. As mentioned earlier, the encryption module depends on a block cipher algorithm such as DES or AES.
5. The super block is represented as an N bit integer. For every channel the remainder of the integer resulting from the previous step will be calculated with respect to the modulus associated with the channel and
6. The calculated residues will be sent on the corresponding channels as packets of data.

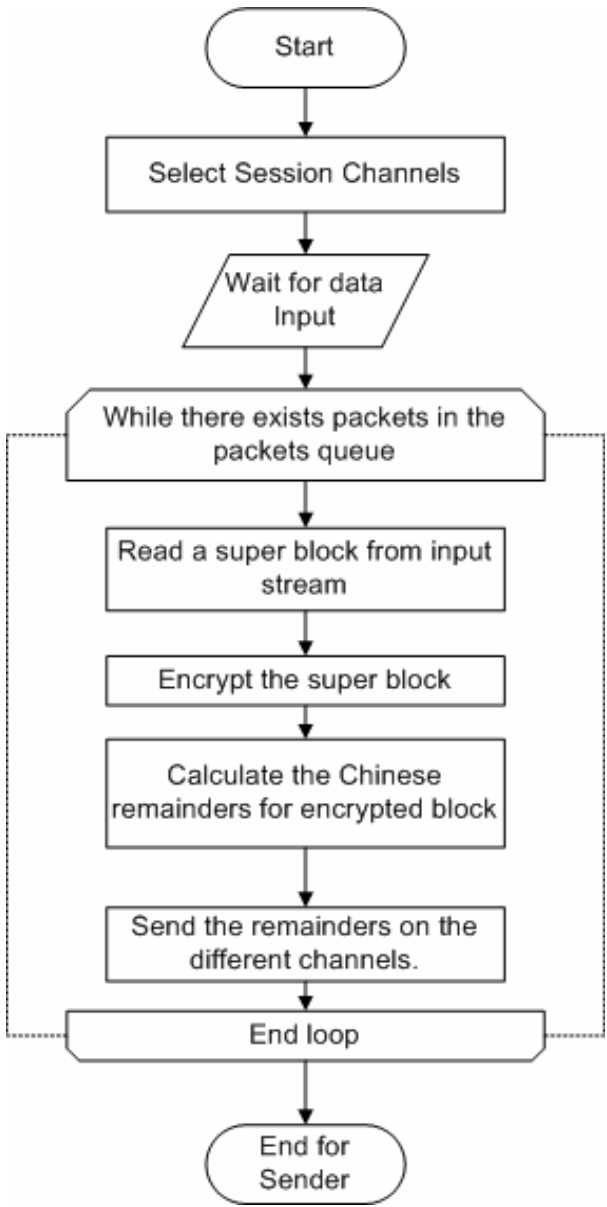

**Figure 4 Sender Flow Chart**

*Receiver Session*

The receiver party receives the data sent over the channels and reconstructs the Message using the following steps:
1. Select channels step, this step is equivalent to the corresponding one on the sender side.
2. Wait for channel input step, this step waits until there are data available on the different channels. This operation is blocking. When the data is available steps three to seven will be repeated until end of transmission.
3. For every channel, an integer r is read that is the remainder sent on the channel.
4. The S integers read from the S channels are used in the inverse Chinese remainder theorem operation to produce an N bit integer.
5. The integer is represented as an N bit super block suitable to be input to the decryption module.
6. The resulting super block is decrypted using the decryption module.
7. The decrypted data is written to the output stream.

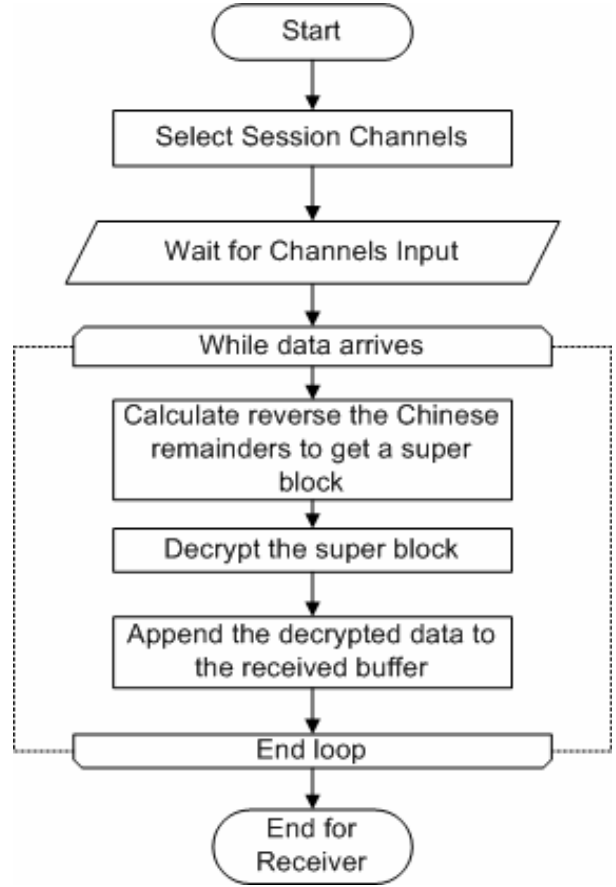

**Figure 5 Receiver Flow chart**

## 4. Security of the proposed scheme

In this section we analyze the security of the proposed scheme. The analysis will start by demonstrating some probabilistic characteristics of the scheme

**Theorem:** The remainders of the CRT are independent.

**Proof:**

Suppose that the output of the encryption module is an N bit value x. Let p and q be two of the S moduli and let Y, Z be random variables denoting the values of $x \bmod p$ and $x \bmod q$ respectively.

Let $y = x \bmod p$ and $z = x \bmod q$. According to the Chinese remainder theorem the ordered pair $(y, z)$ uniquely identifies x in the ranges of values $[0, pq-1]$, $[pq, 2pq-1]$ and so on.

This implies that $P(Y = y \cap Z = z) = 1/pq$.

It is clear that $P(Y = y) = 1/p$ and $P(Z = z) = 1/q$.

The result of the previous steps imply that $P(Y = y \cap Z = z) = P(Y = y)P(Z = z)$ so the two random variables are independent

By generalizing the above proof, it is proved that the remainders of the Chinese Remainder Theorem are independent. Based on this a probability model will be built to describe the output. Let the Random variable C denote the N-Bits output of the Encryption module and define a set of random variables $R_i$ for $0 \leq i < S$ such that $R_i$ denotes the CRT remainder with respect to the modulus $q_i$. The probability distribution of such random variables for $r_i \equiv c \bmod q_i$ can be defined as

$$P(R_0 = r_0, R_1 = r_1..R_{S-1} = r_{S-1}) = \begin{cases} P(C = c) & 0 \leq c < 2^N \\ 0 & otherwise \end{cases}$$

Based on the stated probability relation there are two sets of values that can never appear in the output of the proposed scheme
1. The set of remainders corresponding to the values c such that $2^N \leq c < q_0 q_1 .... q_{S-1}$.
2. For each channel the set of values defined as $x : q_i \leq x < 2^{l_i}$ $l_i = \lceil \log_2 q_i \rceil$.

According to those results the proposed scheme seems to have two points that the adversary can use to compromise the system. According to the concrete security definition [7], the adversary can compromise a security system if he can distinguish it from an equivalent system with truly random output. Based on the demonstrated points in the system we will conclude that theoretically the probability of distinguishability is $P_D = 1 - 2^{N-L}$ it is clear that $P_D \geq 0.5$. But in order to use both of these points the adversary should guess the following:
1. The number of channels S used from within the available channels.
2. The parameters of the pseudorandom generator used to select the S channels from the available A.
3. The relatively prime numbers used in the CRT.

In other words the adversary needs to guess a part of the system key to be able to distinguish it from a random system.

If we assumed that the adversary has an algorithm called the CRT breaker that given a set of channels carrying CRT remainders of encrypted blocks can recover information about the transmitted plain text. The adversary can use CRT Breaker to develop a cipher breaking algorithm as in the following figure to break the underlying encryption scheme given its output.

CipherBreaker($T_C$ /*Cipher text*/)
{
    *Select S relatively prime integers $Q[S]$;*
    *for(int $i = 0; i < S; i++$){*
        $R[i] = T_C \bmod Q[i]$;
    }
    *Return CRTBreaker(Q);*
}

We can conclude from the above algorithm that the proposed scheme security in terms of protecting the plain text is equivalent to the underlying block cipher.

## 5. Conclusion

In this paper a new scheme for transmitting sensitive data is introduced. The proposed scheme was found to have some theoretical issues that prevent it from being secure according to the concrete security definition but was proven to be equivalent to the underlying block cipher.